\documentclass[graybox]{svmult}

\usepackage{mathptmx}       
\usepackage{helvet}         
\usepackage{courier}        
\usepackage{type1cm}        
\usepackage{makeidx}         
\usepackage{graphicx}        
\usepackage{multicol}        
\usepackage[bottom]{footmisc}
\usepackage{listings}

\makeindex             

\definecolor{dkgreen}{rgb}{0,0.6,0}
\definecolor{gray}{rgb}{0.5,0.5,0.5}
\definecolor{mauve}{rgb}{0.58,0,0.82}

\begin{document}

\title*{Reasoning about inter-procedural security requirements in IoT applications}
\author{Mattia Paccamiccio, Leonardo Mostarda}
\institute{Mattia Paccamiccio \at\textbf{Università di Camerino}, Via Andrea d'Accorso 16, 62032 Camerino, \email{mattia.paccamiccio@unicam.it}
\and Leonardo Mostarda \at\textbf{Università di Camerino}, Via Andrea d'Accorso 16, 62032 Camerino, \email{leonardo.mostarda@unicam.it}}
%
%
\maketitle
\abstract{
    \label{chap:abstract}
The  importance of information security dramatically increased and will further grow due to the shape and nature of the modern computing industry. Software is  published at a continuously increasing pace. The Internet of Things and security protocols are two examples of domains that pose a great security challenge, due to how diverse the needs for those software may be, and a generalisation of the capabilities regarding the toolchain necessary for testing is becoming a necessity.
Oftentimes, these software are designed starting from a formal model, which can be verified with appropriate model checkers. These models, though, do not represent the actual implementation, which can deviate from the model and hence certain security properties might not be inherited from the model, or additional issues could be introduced in the implementation.
In this paper we describe a proposal for a novel technique to assess software security properties from LLVM bitcode. We perform various static analyses, such as points-to analysis, call graph and control-flow graph, with the aim of deriving from them an 'accurate enough' formal model of the paths taken by the program, which are then going to be examined via consolidated techniques by matching them against a set of defined rules. 
The proposed workflow then requires further analysis with more precise methods if a rule is violated, in order to assess the actual feasibility of such path(s). This step is required as the analyses performed to derive the model to analyse are over-approximating the behaviour of the software.
}

\section{Introduction}\label{intro}
The verification of software security properties is a critical step when implementing software of any kind. More importantly, in a scenario where we want to assess the logical implementation of a software, we would like to perform, at least in a first instance, the analysis statically because of the following reasons: i) covering all possible environmental conditions can be unfeasible; ii) the code coverage achieved through concrete execution might not be optimal and the poor scaling capabilities of symbolic execution make it unfeasible for this kind of task. The assessment of the logical part of a software is often left to model-checking without considering the code. While the model and its validation can be correct, the implementation can differ from the model, hence there might be the possibility for some properties of the model to be invalidated. 

Other problems arising with certain implementations can be the unavailability of the source code, the toolchain employed to compile it into machine code and/or execute it in a machine with suitable computing power. Dealing with multi-threaded applications increases the complexity of such analyses, as race conditions might happen spuriously if calling order and conditions are not properly enforced. Our approach aims to tackle the detection of violations of inter-procedural requirements in a scalable, fast and static manner.

Consider the following snippet of code:

\begin{lstlisting}[label=lst:socketexample,caption=Example of the usage of sockets in C languase.]
int main(void) {
  struct sockaddr_in servername;
  int sock = socket(PF_INET, SOCK_STREAM, 0);
  if (sock < 0) { exit (EXIT_FAILURE); }
  init_sockaddr (&servername, SERVERHOST, PORT);
  if (0 > connect(sock, (struct sockaddr *)&servername, 
        sizeof(servername))) { exit(EXIT_FAILURE); }
  write_to_server(sock);
  close(sock);
  exit(EXIT_SUCCESS);
}
\end{lstlisting}

Our goal is to check that whenever there is a call to write\_to\_server(), the methods init\_sockaddr(), connect() and close() are also being called, enforcing the order of these calls. In this paper we refer to this type of requirements as \verb|rules|. 

These properties, which in the example above could cause a crash if the program were not to be implemented correctly, could work as well to verify, for instance, that there is no non-encrypted data being sent in certain phases of a hypothetical protocol\cite{blockchainiotautomation}. A higher degree of precision could be obtained by enforcing a specific flow. To further generalise we want to analyse if certain \verb|rules| are always followed in the implementation.

\subsection{Motivation}
\label{Motivation}
Consider the following scenario: the company \verb|XY| has in its core business certain services that are typically used via internet-enabled devices. These devices are oftentimes produced by third-parties, which we will call \verb|ABC|, that take care of the development of the apps that interface with such services via APIs. For the sake of brevity we will assume the example device is a Smart Home Hub device.
XY has the need to validate whether or not the software in the Smart Hub Device is performing the right operations when accessing the APIs. This to prevent having non-handled exceptions visible from the front-end application, crashes or, worse, leakage of sensitive data.
XY has at its disposal the application executable but not its source files. This paper aims to cover this specific use case. 

\subsection{Background: Call graph, Points-to, LLVM}
A call graph is a type of control-flow graph representing calling relationships between the subroutines of a software. The property of a statically computed call graph is to be, typically, an over approximation of the real program's behaviour~\cite{pointeranalysisyannis}.
It is defined as a may-analysis, which means that every calling relationship happening in the program is represented in the graph and possibly some that do not. 
Notice that computing the exact static call graph is an undecidable problem. 

Points-to analysis is a static analysis technique that computes a model of the memory that is used to retrieve what value or set of values a pointer variable will have when de-referenced.
Computing a precise call graph for languages that allow for dynamic dispatching requires a precise points-to analysis and vice-versa. Another way to put it is points-to analysis precision is boosted by a precise call graph computation and vice-versa, because they reduce the sets of possible call relationships and points-to relationships by discarding the impossible ones.

The construction of both call-relationships and points-to relationships have various degrees of sensitivity, which can be used to enhance the precision of the analysis. It can be based on: context (call-site, object), flow.~\cite{pointeranalysisyannis}.

LLVM~\cite{LattnerLLVM2002} is a widely used compiler and tool-chain infrastructure. Its intermediate representation, called LLVM IR, acts as intermediate layer between high-level programming languages and machine code, so that the compiler's duty is to translate the language to LLVM, which is then assembled into the targeted architecture by LLVM itself. The purpose behind this is to simplify the source to binary translation process, by offloading architecture-specific logic from the compiler, task which is instead performed by the LLVM assembler.
\section{Paper contribution}
This paper presents a novel approach for the verification of software security properties, specifically targeting IoT software but potentially extensible to other software domains, such as security protocols. 
Our goal consists in obtaining a inter-procedural control-flow graph, apply transformations to it and using efficient techniques to performs analyses against a set of defined rules.
To reach this goal we need to address the following basic questions: (i) would the approach taken be precise enough for reconstructing the behaviour of the software?; (ii) will using this method provide an advantage in terms of performance compared to other methods of analysis?; (iii) what possible limitations can arise compared to other methodologies?; (iv) how can those be addressed or mitigated?. In order to answer these questions we reviewed the literature on the subject and performed several tests. Such experiments targeted known edge-cases in the analyses we performed. More specifically we assessed the precision for static call-graph generation and points-to and what could be done, if anything, to improve the performance of the analysis.
Our experiments show that our set of approaches could allow for fast and sound analyses, appropriate with the issue being tackled, as also confirmed by the literature.
\section{Related Work}\label{RelatedWork}
In this section we review the state of the art in analyzing and reconstructing the behaviour of a software and give a brief overview on the techniques used, their strengths and weaknesses. 
\subsection{Concrete execution}
Approaches based on concrete execution consist in executing the software in instrumented environments. The way they typically work is by the use of test cases, aiming for maximum code coverage.
As mentioned this approach is accurate when it comes to vulnerability discovery and the reconstruction of a program's behaviour but is not best suited for our scope of analysis. It lacks in terms of performance when computing call relationships and, most importantly, the coverage offered is limited to the paths that are actually executed.

\textbf{afl++}\footnote{https://aflplus.plus , https://github.com/AFLplusplus/AFLplusplus} (american fuzzy lop) is a state of the art fuzzer for concrete executions. It works like a traditional fuzzer: a binary is instrumented and random mutated inputs are fed to the program to explore its states. It also supports LLVM bitcode fuzzing.

\subsection{Dynamic symbolic execution}
Approaches based on symbolic execution consist in statically exploring a binary, representing, ideally, all possible combination of paths it can take as states\cite{ecasymbolic}. Its scalability is poor due to the state-explosion problem which is an open research problem. Methods to mitigate the issue have been proposed and implemented, by selecting promising paths~~\cite{stephens2016driller}, by the means of executing the program backwards~\cite{directedsymexecution,aumaticdiscoverybackwardsbasile}, and merging paths on loops~\cite{veritesting}.

\textbf{Valgrind}~\cite{valgrind} is an instrumentation framework for building dynamic analysis tools. There are Valgrind tools that can automatically detect many memory management and threading bugs, and profile  programs in detail.

\textbf{klee}~\cite{klee} is a symbolic execution tool capable of automatically generating tests that achieve high coverage on a diverse set of complex and environmentally-intensive programs. There are two main components: i) the symbolic virtual machine engine built on top of LLVM,  ii) a POSIX/Linux emulation layer, which also allows to make parts of the operating system environment symbolic.

\subsection{Hybrid approaches}
The capabilities of symbolic execution and dynamic execution can be combined in order to mitigate each techniques' limitations. Dynamic execution's main weakness is the lack of semantic insights inside the program's reasoning and is prone to getting stuck. This can be mitigated with symbolic execution, which offloads dynamic execution by solving complex constraints, while the dynamic execution engine is fast at exploring the 'trivial' parts of a program and does not suffer from state explosion.

\textbf{angr}~\cite{shoshitaishvili2016state} is a programmable binary analysis toolkit that performs dynamic symbolic execution and various static analyses on binaries. It allows for \verb|concolic| execution: a virtual machine emulates the architecture of the loaded binary and there is a concrete execution driven by the symbolic exploration engine. This allows to retrieve concrete data from symbolic constraints, allowing for SMT solver offloading and state thinning.

\textbf{driller}~\cite{stephens2016driller}, based on angr, is a successful example of a symbolic execution engine combined with the traditional fuzzer, afl, with the benefits mentioned above. `

\subsection{Static Analysis}
Approaches based on static analysis work by reasoning about the software without executing it.
Applications of the usage of static analysis techniques are: i) disassemblers, ii) linters, iii) data-flow analyses, iv) model checking. 

\textbf{Ddisasm}~\cite{ddisasm} is a fast and accurate disassembler. It implements multiple static analyses and heuristics in a combined Datalog implementation. Ddisasm is, at the moment, the best-performing disassembler aimed at reassembling binaries.

\textbf{cclyzer}~\cite{cclyzerptsto} is a tool for statically analyzing LLVM bitcode. It is built on Datalog and works by querying a parsed version of the LLVM bitcode. It then performs points-to analysis and call-graph construction on the facts generated. It is capable of yielding highly precise analyses. It supports call-site sensitivity.

\textbf{RetDec}~\cite{retdec} is a retargetable decompiler based on LLVM. Its main feature is the conversion of machine code to LLVM bitcode. It supports reconstruction of functions, types, and high-level constructs. It is capable of generating call graphs and control-flow graphs.

\textbf{McSema}\footnote{https://github.com/lifting-bits/mcsema} offers very similar functionalities and can be applied to the same scope as RetDec. Its main difference from it is to use external software to deal with control-flow graph recovery (a step of paramount importance for disassembling).  

\section{Proposed approach}\label{OurApproach}
\begin{figure}
    \centering
    \includegraphics[width=\linewidth]{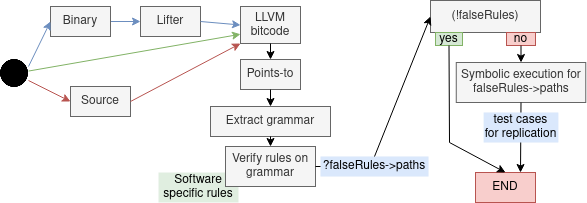}
    \caption{Our approach.}
    \label{fig:llvmcheck_drawio}
\end{figure}

Our approach can be summarised by the diagram in figure \ref{fig:llvmcheck_drawio} and will be further explained in the following sections.

\subsection{Call relationships}
\label{subsec:callrelationships}
We chose to base our workflow on LLVM, which allows to be quasi-agnostic to binary, source or LLVM bitcode as a starting point, as source can be compiled to LLVM and binary can be lifted to LLVM bitcode.
The LLVM bitcode will then be used to perform points-to analysis with online call-graph construction, so to get a accurate inter-procedural and infra-procedural view of the software.
We would obtain a model which would then be validated against our set of rules.



This could be formalised as follows:

\begin{equation}\label{callrel}
        \forall f, \forall c ( \,f ) \, \rightarrow \exists \varphi \in \overline{\rm f}( \,c ( \,f ) \,) \, \cup\,\overline{\rm f} :\varphi\uparrow
\end{equation}

The previous formula is read as follows: for each function $f$ and $c(f)$ ($c$ calls function $f$), there exists a set $\varphi$ that is a subset of $\overline{\rm f}(c(f))$ (set of functions that call functions calling $f$) and $\overline{\rm f}$ (that do not call $f$ and $c$) so that $\varphi$ is converging (up arrow). The analysis, as introduced here, only reasons about call-relationships with no considerations regarding flow. We call this property \verb|converge|.
Our goal for the scope of this paper has been to assess whether or not the model of tested software can be tested against this rules and similar ones. 

Consider the following example rule that encodes the fact that the call order is connect, followed by write\_to\_server, followed by close:

\begin{lstlisting}
rule {  properties[] = ["converge","order"], 
        f[] = ["connect", "write_to_server", "close"]; }
\end{lstlisting}

What we want to verify is that for each of the functions specified in f[] the property is ensured. The property \verb|converge| is explained in the equation \ref{callrel}. The property \verb|order| means: the functions in f[] are executed in the specified order. 

\subsection{Program behaviour extraction}
\label{programbehaviourextraction}
For complexity reasons we deem necessary to further simplify the model we analyse, instead of working directly with control-flow graphs. As mentioned in \cite{atomicityviolation} we can transform control-flow graphs into context-free grammars. This allows for very efficient exploration and further simplification of the model without compromising soundness and accuracy for the intended scope.

\subsection{Validation and refinement}
Once the behaviour has been extracted and matched against our set of rules, we obtain the violations, if present.
Since we are working in an over-approximated realm we propose to apply selective symbolic execution to the paths that failed the validation in order to retrieve if: i) if such paths can exist; ii) test cases to replicate the failure; iii) information to help debugging by locating the failure.
\section{Experiments}
\label{chap:ExResults}
To evaluate if our methodology is feasible we tested our workflow on edge-cases that make points-to analysis necessary to obtain an accurate call-graph.
These tests included: i) utilisation of pointers to data, ii) utilisation of calls and branched calls to function pointers.
The following snippets of code in C (compiled to LLVM) are the examples we performed validations on:
\begin{lstlisting}[label=lst:branchedfunptr,caption="branched\_funptr" snippet.]
int main() 
{   int in = 3;
    void (*one)();
    if (in > 5) {
      one = &first;
    } else {
      one = &second;
    }
    (*one)();
    return 0; }
\end{lstlisting}
\begin{lstlisting}[label=lst:branchedfunptrorder,caption="branched\_funptr\_order" snippet.]
int main() 
{   int in = 3;
    void (*one)(), void (*two)();
    if (in > 5) {
      one = &first;
      two = &second;
    } else {
      one = &second;
      two = &first;
    }
    (*one)();
    (*two)();
    return 0; }
\end{lstlisting}
\begin{lstlisting}[label=lst:firstsecond,caption=Methods "first" and "second".]
void first(){ puts("First"); } void second(){ puts("Second"); }
\end{lstlisting}
For the sake of brevity we included only the snippets involving branched function pointers.
At the time of writing we did not account for control-flow in our experiments but just on pure call relationships, as control-flow recovery can be obtained once the inter-procedural relationships are accurate.
Additional analyses on GNU binaries have been performed to test how fast the analysis would have been on more sizeable software, considering performances only and not accuracy. The validation of the rule formalised in \ref{subsec:callrelationships} is left behind because the performance load is negligible compared to the one given by computing the call-graph and points-to relationships.

\subsection{Experimental Results}
We obtained positive results using Cclyzer\cite{cclyzerptsto} when performing points-to and call-graph generation on LLVM bitcode. We also noted how the analysis becomes slower when working with higher call-site sensitivities. 
We also tested RetDec\cite{retdec} and McSema\footnote{https://github.com/lifting-bits/mcsema} as lifters, with the latter being the most accurate one. We did not perform more detailed experiments on this aspect at the time of writing because we assumed that we can have correct LLVM bitcode. Table \ref{tbl:performancecclyzer} summarises the experimental results obtained with CClyzer. Such results show performance (time taken) only, based on sensitivity and complexity (file size). 

\begin{table}[]
    \centering
    \begin{tabular}{ |p{3cm}|p{2cm}|p{1.3cm}|p{1.3cm}|  }
     \hline
     Target & Sensitivity & Size (KB) & Time (s)\\
     \hline
     fptr (snippets)            & call-site-2    &  2.5   & 6.39\\
     br\_fptr (snippets)        & call-site-2    &  2.8   & 6.12\\
     fptr\_order (snippets)     & call-site-2    &  2.6   & 6.33\\
     br\_fptr\_order (snippets) & call-site-2    &  2.8   & 6.24\\
     \hline
     dd (coreutils)             & insensitive & 273     & 14.27\\
     dd (coreutils)             & call-site-2 & 273     & 16.23\\
     ls (coreutils)             & insensitive & 612.3   & 22.57\\
     ls (coreutils)             & call-site-2 & 612.3   & 27.38\\
     cp (coreutils)             & insensitive & 733     & 28.50\\
     cp (coreutils)             & call-site-2 & 733     & 31.49\\
     \hline
    \end{tabular}
    \caption{CClyzer performance tests for points-to analysis and call-graph construction. "insensitive" analyses do not take into account call-site, "call-site-2" account for call-site up to depth 2. Among the snippets the words: fptr, br, order mean respectively: call to function pointer, branched and change of order, as shown in \ref{chap:ExResults}. }
    \label{tbl:performancecclyzer}
\end{table}

\section{Conclusion and Future Work}
\label{conclusion}
\label{chap:Conclusions}
The first results of our methodology are positive and allow for basic validations on our test samples. The test samples were designed to target applications where points-to analysis is made necessary and met our expectations, both in precision and performance.
\subsection{Future developments}
Future work includes adding parallel and flow reasoning to the framework, thus extending the application to detection of call-order violations on both single and multi-threaded application, enabling validation of operational order.
This task is made easier by LLVM being a Single Static Assignment, hence having intrinsic flow-sensitivity~\cite{pointeranalysisyannis}. To implement this step we propose to derive context-free grammars from control-flow graphs, as discussed in \ref{programbehaviourextraction} and demonstrated in~\cite{atomicityviolation}, allowing for both call-order enforcement and verification of atomicity properties in parallel software. To narrow down unfeasible paths, we propose to use symbolic execution to generate use-cases targeting the path(s) violating the rule(s) as it is more precise than the previously performed analyses.
%
%
%
\bibliographystyle{elsarticle-num}
\bibliography{cas-refs}

\begin{thebibliography}{10}
\expandafter\ifx\csname url\endcsname\relax
  \def\url#1{\texttt{#1}}\fi
\expandafter\ifx\csname urlprefix\endcsname\relax\def\urlprefix{URL }\fi
\expandafter\ifx\csname href\endcsname\relax
  \def\href#1#2{#2} \def\path#1{#1}\fi

\bibitem{blockchainiotautomation}
R.~Sekaran, R.~Patan, A.~Raveendran, F.~Al{-}Turjman, M.~Ramachandran,
  L.~Mostarda, Survival study on blockchain based 6g-enabled mobile edge
  computation for iot automation, {IEEE} Access 8 (2020).

\bibitem{pointeranalysisyannis}
S.~Yannis, B.~George, Pointer analysis, Found. Trends Program. Lang. 2~(1)
  (2015) 1--69.

\bibitem{LattnerLLVM2002}
C.~Lattner, {LLVM: An Infrastructure for Multi-Stage Optimization}, Master's
  thesis, {Computer Science Dept., University of Illinois at Urbana-Champaign},
  Urbana, IL (Dec 2002).

\bibitem{ecasymbolic}
C.~Vannucchi, M.~Diamanti, G.~Mazzante, D.~Cacciagrano, R.~Culmone,
  N.~Gorogiannis, L.~Mostarda, F.~Raimondi, Symbolic verification of
  event-condition-action rules in intelligent environments, J. Reliab. Intell.
  Environ. 3~(2) (2017).

\bibitem{stephens2016driller}
N.~Stephens, J.~Grosen, C.~Salls, A.~Dutcher, R.~Wang, J.~Corbetta,
  Y.~Shoshitaishvili, C.~Kruegel, G.~Vigna, Driller: Augmenting fuzzing through
  selective symbolic execution (2016).

\bibitem{directedsymexecution}
K.~Ma, Y.~P. Khoo, J.~S. Foster, M.~Hicks, Directed symbolic execution, in:
  E.~Yahav (Ed.), Static Analysis - 18th International Symposium, {SAS} 2011,
  2011. Proceedings, Vol. 6887 of Lecture Notes in Computer Science, Springer,
  2011, pp. 95--111.

\bibitem{aumaticdiscoverybackwardsbasile}
C.~Basile, D.~Canavese, J.~d'Annoville, B.~D. Sutter, F.~Valenza, Automatic
  discovery of software attacks via backward reasoning, in: P.~Falcarin,
  B.~Wyseur (Eds.), 1st {IEEE/ACM} International Workshop on Software
  Protection, {SPRO} 2015, {IEEE} Computer Society, 2015, pp. 52--58.

\bibitem{veritesting}
T.~Avgerinos, A.~Rebert, S.~K. Cha, D.~Brumley, Enhancing symbolic execution
  with veritesting, in: P.~Jalote, L.~C. Briand, A.~van~der Hoek (Eds.), 36th
  International Conference on Software Engineering, {ICSE} '14, 2014, {ACM},
  2014, pp. 1083--1094.

\bibitem{valgrind}
N.~Nethercote, J.~Seward, Valgrind: a framework for heavyweight dynamic binary
  instrumentation, in: J.~Ferrante, K.~S. McKinley (Eds.), Proceedings of the
  {ACM} {SIGPLAN} 2007 Conference on Programming Language Design and
  Implementation, 2007, {ACM}, 2007, pp. 89--100.

\bibitem{klee}
C.~Cadar, D.~Dunbar, D.~R. Engler, {KLEE:} unassisted and automatic generation
  of high-coverage tests for complex systems programs, in: R.~Draves, R.~van
  Renesse (Eds.), 8th {USENIX} Symposium on Operating Systems Design and
  Implementation, {OSDI} 2008, {USENIX} Association, 2008, pp. 209--224.

\bibitem{shoshitaishvili2016state}
Y.~Shoshitaishvili, R.~Wang, C.~Salls, N.~Stephens, M.~Polino, A.~Dutcher,
  J.~Grosen, S.~Feng, C.~Hauser, C.~Kruegel, G.~Vigna, Sok: (state of) the art
  of war: Offensive techniques in binary analysis (2016).

\bibitem{ddisasm}
A.~Flores{-}Montoya, E.~M. Schulte, Datalog disassembly, CoRR abs/1906.03969
  (2019).

\bibitem{cclyzerptsto}
G.~Balatsouras, Y.~Smaragdakis, Structure-sensitive points-to analysis for {C}
  and {C++}, in: X.~Rival (Ed.), Static Analysis - 23rd International
  Symposium, {SAS} 2016, Proceedings, Vol. 9837 of Lecture Notes in Computer
  Science, Springer, 2016, pp. 84--104.

\bibitem{retdec}
J.~K\v{r}oustek, P.~Matula, P.~Zemek, Retdec: An open-source machine-code
  decompiler, [talk], presented at Botconf 2017, Montpellier, FR (December
  2017).

\bibitem{atomicityviolation}
D.~G. Sousa, R.~J. Dias, C.~Ferreira, J.~Louren{\c{c}}o, Preventing atomicity
  violations with contracts, CoRR abs/1505.02951 (2015).

\end{thebibliography}

\end{document}